\def\USEACHEMSO{1} 

\if\USEACHEMSO1
\documentclass[journal=jacsat,manuscript=communication,email=true,]{achemso}
\else
\documentclass[aip,notitlepage,
tightenlines,
twocolumn,
citeautoscript, reprint,floatfix]{revtex4-2} 
\fi

\usepackage[utf8]{inputenc}
\usepackage[T1]{fontenc} 
\usepackage[version=3]{mhchem}
\usepackage{hyperref}
\usepackage{units}
\usepackage{booktabs}
\usepackage{siunitx}
\usepackage{longtable}
\usepackage{xspace}
\usepackage{verbatim}

\usepackage{graphicx} \usepackage[textwidth=2.2cm]{todonotes}
\newcommand{\angstrom}{\textup{\AA}} \newcommand{\new}{}

\if\USEACHEMSO1
\fi

\if\USEACHEMSO0
\begin{document}
\fi

\title{The chromium dimer: closing a chapter of quantum chemistry}

\author{Henrik R.~Larsson}
\affiliation{Division of Chemistry and Chemical Engineering, California Institute of Technology, Pasadena, CA 91125, USA}
\if\USEACHEMSO1
\alsoaffiliation{Department of Chemistry and Biochemistry, University of California, Merced, CA 95343, USA}
\else
\affiliation{Department of Chemistry and Biochemistry, University of California, Merced, CA 95343, USA}
\fi
\email{larsson22\_cr2 [a t] larsson-research. de}
\author{Huanchen Zhai}
\affiliation{Division of Chemistry and Chemical Engineering, California Institute of Technology, Pasadena, CA 91125, USA}
\author{C. J. Umrigar}
\affiliation{Laboratory of Atomic and Solid State Physics, Cornell University, Ithaca, NY 14853, USA}
\author{Garnet Kin-Lic~Chan}
\email{garnetc [a t] caltech . edu}
\affiliation{Division of Chemistry and Chemical Engineering, California Institute of Technology, Pasadena, CA 91125, USA}

\if\USEACHEMSO1
\begin{document}
\fi

\if\USEACHEMSO1
\fi

\begin{abstract} 
The complex electronic structure and unusual potential energy curve of the chromium dimer have fascinated scientists for decades, with agreement between theory and experiment so far elusive. 
Here, we present a new ab initio simulation of the potential energy curve and vibrational spectrum that significantly improves on all earlier estimates. Our data support a shift in earlier experimental assignments of a cluster of vibrational frequencies by one quantum number. 
The new vibrational assignment yields an experimentally derived potential energy curve in quantitative agreement with theory across all bond lengths and across all measured frequencies.
By solving this long-standing problem, 
our results raise the possibility of quantitative quantum chemical modeling of transition metal clusters with spectroscopic accuracy.
\end{abstract}
\maketitle

\maketitle
Transition metal chemistry plays a pivotal role in catalysis, biochemistry, and the energy sciences, but the complex electronic structure of the $d$-shells challenges our modeling and understanding of such processes~\cite{kurashige_2013,sharma_2014b,li2019electronic}. 
Amongst the most complicated of small transition metal molecules is the chromium dimer, 
often described as a grand challenge problem of small molecule quantum chemistry,
and whose unusual bonding and potential energy curve (PEC) has puzzled scientists for decades~\cite{Bimetal1977klotzbuecher,cr21978cotton,cr21981goodgame,THEORETICAL1983walch,cr21985goodgame,cr21983bondybey,cr21976anderson,casey_1991,cr2PEC1993casey,Cr21994bauschlicher,thomas_1999,roos_2003,cr22007roos,SplitGAS2013limanni}.
\new{Addressing this is relevant for other dichromium compounds, as well as for other compounds where multiple metal-metal bonds and spin coupling appear.\cite{Bimetal1977klotzbuecher,cr21978cotton,cr21985goodgame,cr21976anderson,kurashige_2013,sharma_2014b,li2019electronic,Quantum2006bryndaa,cr22007roos,
wagner2009ultrashort,Orbital2019stein,Interpretation2021chalupsky} }
The \ce{Cr-Cr} bond in the bare dimer is a formal sextuple bond, and when complexed with ligands, was the first example of a quintuple bond~\cite{cr2Quint2005nguyen,Quantum2006bryndaa}.
Although the formal bond order is high, the \new{PEC} inferred from photoelectron spectroscopy indicates a short and weak bond with a narrow minimum around $\unit[1.68]{\angstrom}$, and an extended shelf at around 
$\unit[2.5]{\angstrom}$~\cite{cr2PEC1993casey}. The curve takes this form because the \ce{Cr}
4s and 3d atomic orbitals are of very different size, with the minimum corresponding mostly to 3d orbital interactions and the shelf to 4s orbital interactions. Beyond this picture, a quantitative understanding  remains lacking. In particular, theoretical predictions of the binding curve deviate substantially from the experimentally derived curve, as well as from each other, while the experimental curve is uncertain at longer bond lengths. Here we show that a combined analysis from new numerical simulations using state-of-the-art quantum chemistry methods, together with  existing experimental data, yields a definitive picture of the \ce{Cr2} \new{PEC}. In particular, our work suggests a reassignment of the vibrational subbands in the shelf region, bringing theory and experiment finally into quantitative agreement.

The complex electronic structure arises from the interplay of two types of electron correlation. First, there is the spin-coupling of the 12 valence electrons in the 3d and 4s Cr atoms shells; this is termed \new{static correlation}. Because the many-electron wavefunction of the valence electrons is not well captured by a single \new{determinant}, we refer to the electronic structure as multireference. Second, a large basis is needed to capture excitations involving non-valence orbitals; for example, the formation of the 3d-3d bonds requires the 3p electrons to move out of the same spatial region by exciting to higher lying orbitals; such effects are referred to as  dynamic correlation. The problem is computationally challenging because both the static and dynamic correlation must be computed sufficiently well even for a qualitatively reasonable description. For example, the valence complete active space self-consistent-field method (which treats the valence static correlation exactly but neglects the dynamic correlation) does not yield a minimum near the equilibrium bond length,\new{\cite{THEORETICAL1983walch}}
while the gold-standard treatment of dynamical correlation, coupled cluster singles, doubles and perturbative triples (CCSD(T)) also does not yield a reasonable bond length, nor does it display a shelf region.\new{~\cite{Cr21994bauschlicher} }

\begin{figure*}[!tbp]
\includegraphics[width=1.5\columnwidth]{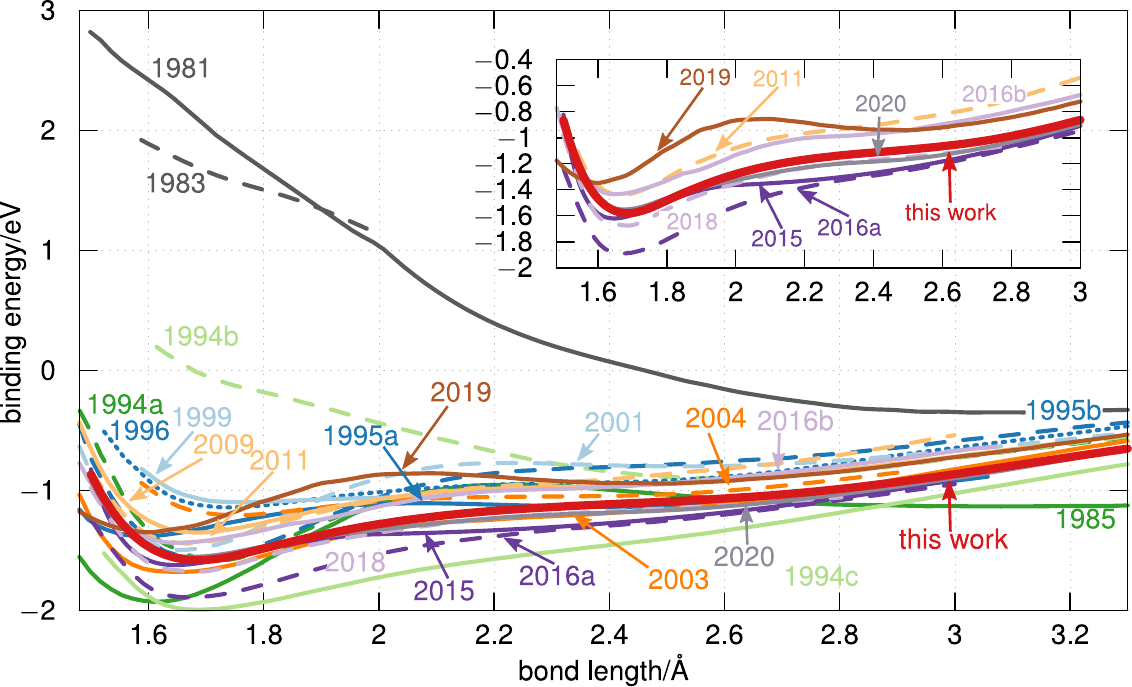}
\caption{Some of the simulated potential energy curves (PECs) of the chromium dimer \new{available} in the literature\new{, labeled by year}. 
The red curve marks \new{this work}. The inset shows selected PECs from 2011 onwards. }
  \label{fig:cr2_lit}
\end{figure*}

\begin{figure*}[!tbp]
\includegraphics[width=1.5\columnwidth]{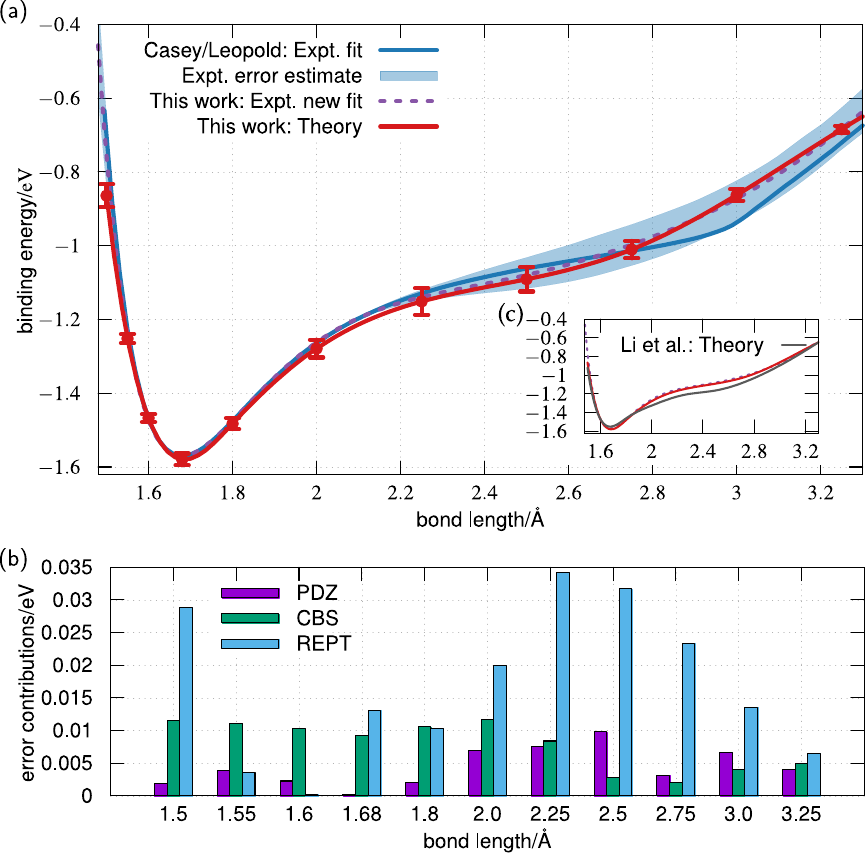}
\caption{
    Theoretical and experimental potential energy curves (PECs) of the chromium dimer. 
    (a) Blue: experimental PEC from Ref.~\citenum{cr2PEC1993casey}. The blue-shaded area estimates
    the uncertainty from the experimental PEC fit. 
    Purple: new experimental PEC from vibrational assignment in this work. Red: computed PEC with error estimates.
    (b) Estimated simulation errors. $\epsilon_\text{PDZ}$: error of ``exact'' estimate of the \new{cc-pVDZ-DK} basis curve. $\epsilon_\text{CBS}$: \new{error} in the complete basis-set extrapolation. $\epsilon_\text{REPT}$: \new{error} of the dynamic correlation correction. (c) New theoretical and experimental PEC compared with the next best PEC in the literature from Li et al. \new{(gray curve).\cite{li2020CrSHCI}}
    }
  \label{fig:cr2_pes}
\end{figure*}

\autoref{fig:cr2_lit} shows \new{curves} from calculations over many years; the lack of consensus is striking. Even when limited to studies from the last decade, there is a spread of over $\unit[0.6]{eV}$ in the predicted binding energy and $\unit[0.2-0.34]{eV}$ across the whole curve. Experimentally, while multiple techniques have shed light on the spectroscopic constants,
information on the full PEC comes from a photoelectron spectroscopy study of $\ce{Cr2-}$~\cite{casey_1991,cr2PEC1993casey}. This measured 29 vibrationally resolved transitions, and by assigning these to specific vibrational quantum numbers $v$,
a PEC was derived using the Rydberg-Klein-Rees method. However, the assignments above $v = 9$ are uncertain, in particular, the assignment of the starting quantum number $v_\text{prog}$ of a high-lying 20 member vibrational progression. Together with non-uniqueness in the PEC fit, this leads to considerable uncertainty in the experimental PEC. This is shown
by the shaded region of \autoref{fig:cr2_pes}a, which shows the range of experimental PEC arising from different assignments ($v_\text{prog}=21-25$), all of which match the observed vibrational levels within their experimental uncertainty; in the shelf-region the uncertainty is over $\unit[0.1]{eV}$. (Further details in the SI).

\new{To compute a more accurate PEC,}
we will \new{employ a composite method} starting from the scalar relativistic \new{``exact two-component''} (X2C) Hamiltonian,\cite{Exact2012peng,Quasirelativistic2005kutzelnigg}
\new{which is based on two contributions.}
The first \new{contribution} estimates  the exact chromium dimer PEC in a moderate basis (Dunning's cc-pVDZ-DK basis, here dubbed PDZ~\cite{dunningBas2006balabanov}, with a frozen neon core, correlating 28 electrons in 76 orbitals). 
For this, we use data from very large ab initio density matrix renormalization group (DMRG) calculations (using up to bond dimension 28000 and SU(2) symmetry), together with selected heat-bath configuration interaction (SHCI) data computed \new{earlier} by one of us.\cite{li2020CrSHCI} The second \new{contribution} targets the remaining dynamic \new{correlation.} For this, we use data computed from multi-reference perturbation theory (using an efficient formulation of the restraining the excitation-order Hamiltonian~\cite{fink2009MRrept,sharma2015REPT} \new{within} the language of matrix product state perturbation theory~\cite{sharma2014Hylleraas,sharma2015REPT,bigSite2021larsson}, starting from the 12 electron, 12 orbital valence complete active space) computed using cc-pV$N$Z-DK basis sets up to quintuple zeta, as well as unrestricted CCSD(T) data~\cite{li2020CrSHCI}. This dynamical correlation correction is then extrapolated to the basis set limit. 
We perform the simulations with the PySCF and block2 program packages~\cite{PYSCF1,PYSCF2,zhai2021Low,bigSite2021larsson}.

It is important to estimate the error in these various contributions. For the PDZ curve, as DMRG and SHCI provide independent extrapolations to numerical exactness with similar confidence, we take the average of the DMRG and SHCI data as the \new{curve}, with half the difference as the \new{error} ($\epsilon_\text{PDZ}$). For the dynamic correlation correction, the CCSD(T) data is expected to be less accurate than the REPT data, due to the multireference nature of the correlation. Thus we only use the REPT \new{data,} and use half the difference from CCSD(T) as the error ($\epsilon_\text{REPT}$). Finally, the basis set error is estimated as the standard deviation of the complete basis set \new{(CBS)} extrapolation fit \new{($\epsilon_\text{CBS}$).}
Taking these three error contributions as independent, the total error is then the square root of the quadratic sum.
\new{Note} that it is difficult to assert the statistical significance of these error estimates, however, they provide a useful measure of accuracy.

Using the new theoretical PEC, we compute the detailed vibrational spectrum by solving \new{the} vibrational Schr\"odinger equation. We  use this to then reassign the measured experimental peaks from the photoelectron spectrum~\cite{cr2PEC1993casey}, and with these assignments, solve the inverse Schr\"odinger equation to derive a new experimental PEC.

We show the computed \new{PEC} in ~\autoref{fig:cr2_pes}a.
The accompanying error estimates for the {PDZ} curve, the dynamic correlation correction,  and the \new{CBS} extrapolation are shown in~\autoref{fig:cr2_pes}b. 
$\epsilon_\text{PDZ}$ is quite small $<\unit[0.01]{eV}$, demonstrating remarkable agreement in the ``exact'' PDZ energies from DMRG and SHCI.  \new{$\epsilon_\text{CBS}$} is also small $<\unit[0.012]{eV}$. The largest error is from the \new{dynamic correlation}
which is as large as $\sim \unit[0.034]{eV}$ at $\unit[\sim2.0]-\unit[2.25]{\angstrom}$. As discussed in the SI, this likely reflects the
poor performance of the CCSD(T) method used to estimate the error, 
and thus this large \new{error} is a conservative estimate (other ways \new{of estimation} given in the SI).
Overall, we find good agreement with the existing experimental PEC, and one that is significantly improved over all previous computations in the literature (the next best match is shown in~\autoref{fig:cr2_pes}c, which has substantial disagreement in the shelf region). Unlike some earlier predictions, no double minimum is observed. The largest uncertainties in the theoretical curve lie outside of the region of the PEC with large experimental uncertainty; we return to this point below.

\begin{figure*}[!tbp]
\includegraphics[width=1.5\columnwidth]{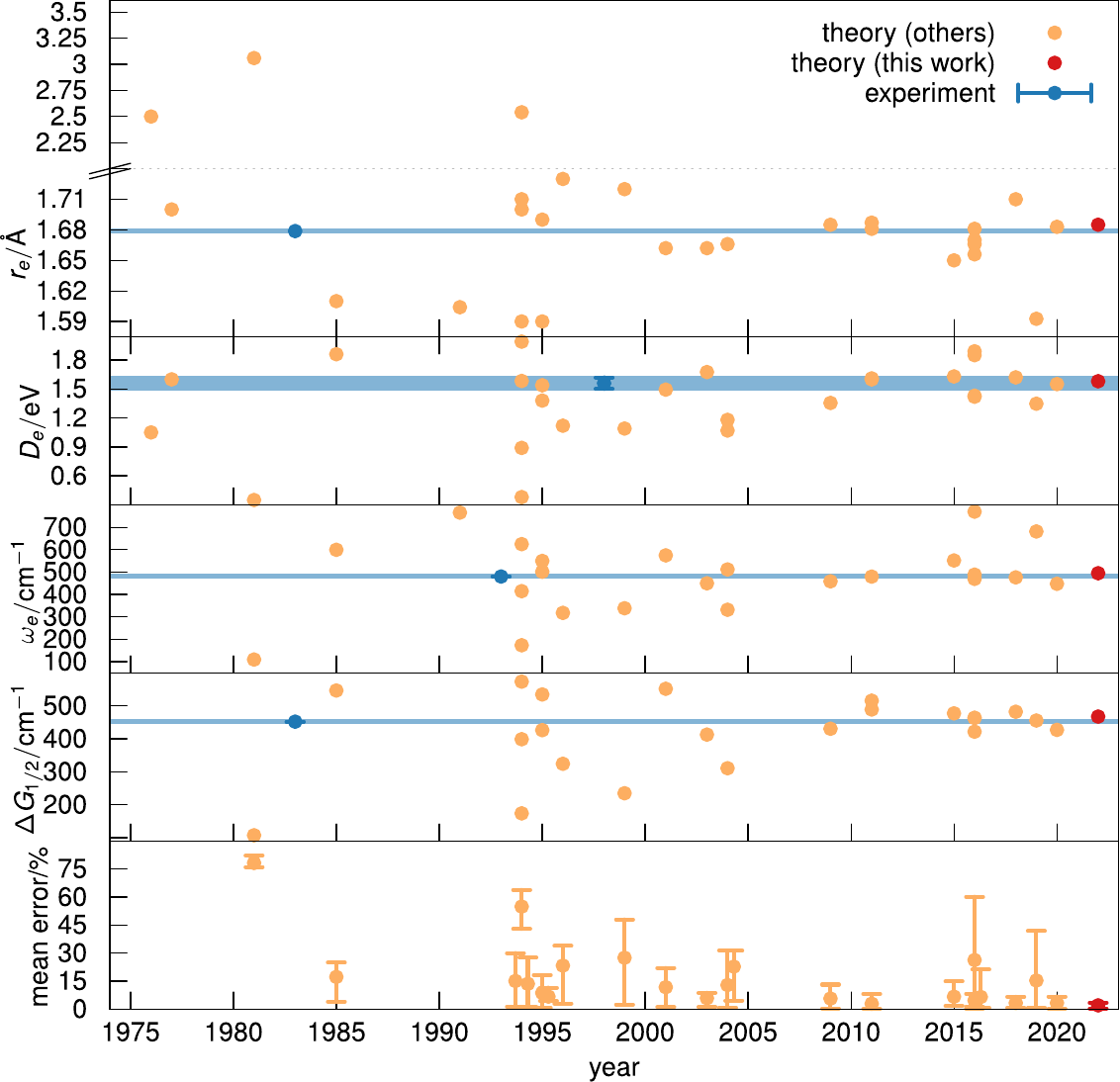}
    \caption{
    Simulated spectroscopic constants of \ce{Cr2} over time: 
    $r_e$ (equilibrium bond length), $D_e$ (well-depth), $\omega_e$ (harmonic frequency) and $\Delta G_{1/2}$ (fundamental frequency). 
    Blue: most recent experimental result; shaded area: experimental uncertainty \new{(that of $r_e$ is not reported)}. The lowest panel shows the evolution of the average absolute percentage \new{error} (in case all four constants are available; error bar shows min/max error). 
\new{Further data} is shown in the SI (Table S5).
    }
  \label{fig:cr2_constants}
\end{figure*}

\autoref{fig:cr2_constants} shows  spectroscopic constants derived from the current and earlier PECs, compared to \new{experiment}. We find very good agreement with experiment; the improvement in theoretical predictions over time is shown in the lowest panel. Note, however, that the spectroscopic constants only measure the quality of the \new{PEC} only near the minimum. In fact, the other studies with spectroscopic constants with small mean error \new{are} associated with \new{PECs} of poor overall shape (see~\autoref{fig:cr2_lit}), reflecting a poor description of the full vibrational spectrum.

\begin{figure*}
\centering
  \includegraphics[width=1.2\columnwidth]{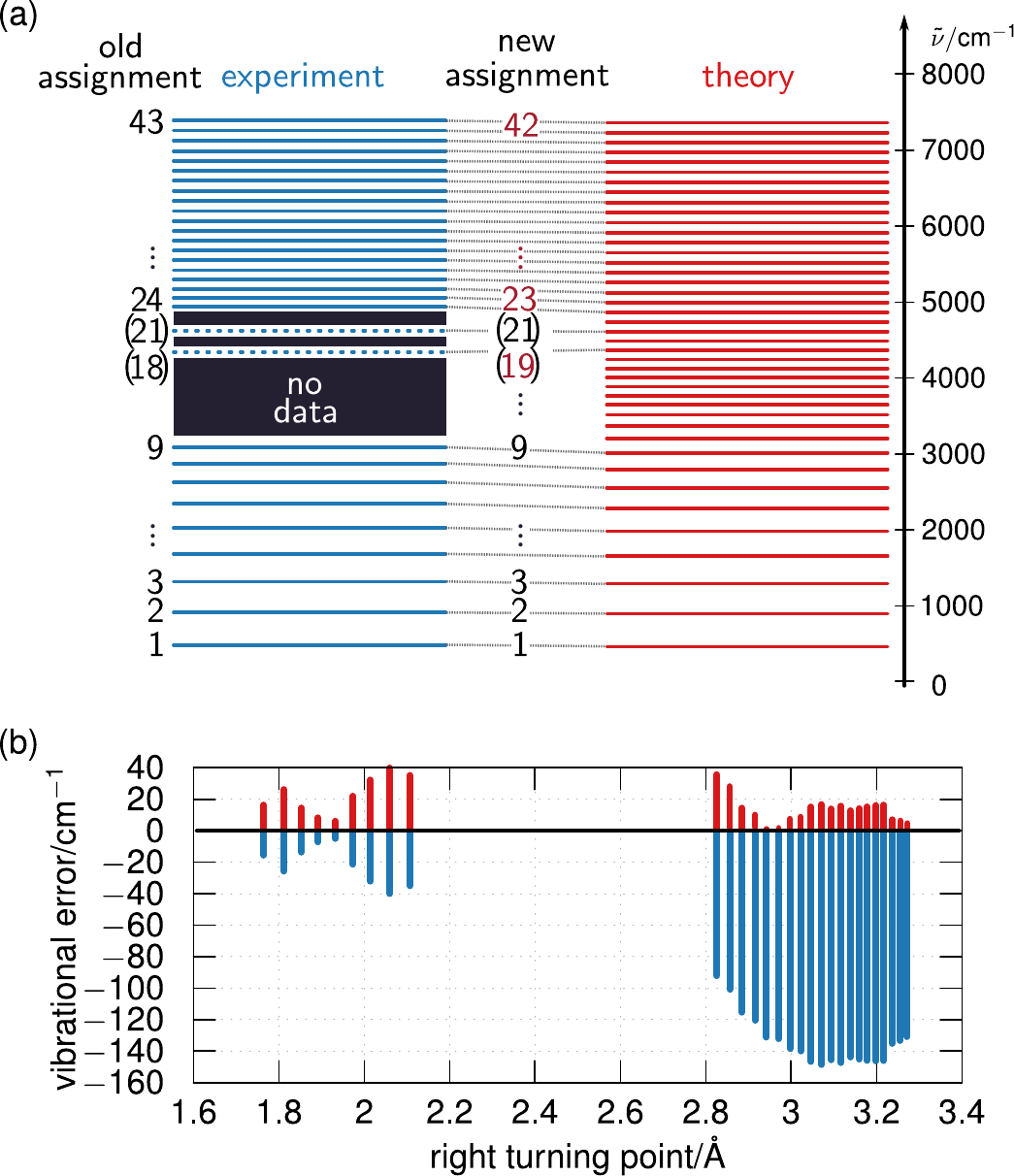}
  \caption{Vibrational ladder (a): Experimental (left) vibrational frequencies compared with the simulated frequencies (right). The experimental data lacks values between $\unit[\sim 3250]{cm^{-1}}$ and \mbox{$\unit[\sim 4750]{cm^{-1}}$}; the original assignment of  frequencies with quantum numbers larger than 9 is not fully certain.
  (Note that the two measured frequencies labeled in \new{brackets} have not definitively been associated with the \ce{Cr2} ground state, see Ref.~\citenum{cr2PEC1993casey}). 
  The simulated frequencies enable a new \new{assignment}, shown in red. 
  (b): Vibrational error of the new (old) assignment shown as positive (negative) values, as function of the right classical turning point of each state (based on our new experimental PEC). 
  For the 9 lowest vibrational levels, assiociated with turning points up to $\unit[2.1]{\angstrom}$, the assignment does not change and the errors are identical. For the 20 higher-lying levels, the old (new) assignment uses $v_\text{prog}=24 (23)$.
}
  \label{fig:freqs}
\end{figure*}

We now more carefully examine the shelf region of the \new{PEC}, where there is the largest deviation from the experimentally derived curve. 
Ref.~\citenum{cr2PEC1993casey} contains a progression of 20 vibrational levels starting from $\unit[4880]{cm^{-1}}$, with a spacing of approximately $\unit[128]{cm^{-1}}$.  Casey and Leopold tried various assignments, ultimately assigning the first frequency of this cluster to $v_\text{prog}=24$. The vibrational frequencies for our theoretical \new{PEC} are shown in ~\autoref{fig:freqs}a. 
With the old assignment of $v_\text{prog}$, our simulated frequencies of this cluster consistently disagree with the experimental result by approximately a single energy quantum in the shoulder region of the \new{PEC} (\autoref{fig:freqs}b). This is surprising given the small theoretical estimate of the error in this region, and suggests that we should simply change the assignment from \new{$v_\text{prog}=24$ to $v_\text{prog}=23$}. This reduces the root mean square deviation (RMSD) from $\unit[113]{cm^{-1}}$ to $\unit[19]{cm^{-1}}$. The largest discrepancy now occurs for one of the lower states ($v=7$), consistent with the region of largest uncertainty in the theoretical calculations, between $\unit[2.0-2.25]\angstrom$, (see \autoref{fig:freqs}b). 
Notably, the experimental vibrational frequencies have a RMSD of $\unit[16]{cm^{-1}}$,  similar to the RMSD of the computed frequencies with the new assignment. (See Table S4 in the SI). \new{While theory does not allow for a statistical estimate of certainty, the quantitative agreement between the theoretical and experimental vibrational frequencies across the measured peaks is striking and is our main result. 
As a consequence, solving the inverse Schr\"odinger equation 
with the suggested new assignment of $v_\text{prog}$ 
leads to a revised estimate of the experimental PEC shown in \autoref{fig:cr2_pes}a.
The revised PEC demonstrates an excellent match between theory and experiment.}

The computational prediction of the ground-state \new{PEC} of a diatomic that is quantitatively consistent with experiment might seem to be a standard task, but in the case of the chromium dimer it has been a challenge for decades. Our work shows that this goal can finally be achieved; as one metric, the average error in the vibrational spectrum computed from the theoretical \new{PEC} is now comparable to the average uncertainty of the vibrational peaks measured in experiment. While this arguably brings to a close a storied problem of computational quantum chemistry, it opens the door to many others, in particular, the applications of the theoretical techniques discussed here \new{not only to other complex multiple metal-metal bonded species, but more generally to the quantitative spectroscopic modeling of transition metal clusters.}

\if\USEACHEMSO1
\begin{acknowledgement}
\else
\acknowledgements
\fi
\new{We thank Doreen Leopold and Alec Wodtke for helpful discussions.}
Work by G.K.C.~was supported by the US National Science Foundation (NSF) via grant no.~CHE-2102505. G.K.C.~acknowledges additional support from the Simons Foundation via the Many-Electron Collaboration and the Investigator Award. 
Work by H.R.L., H.Z.~\new{and C.J.U.} was supported by the Air Force Office of Scientific Research, under Award FA9550-18-1-0095. 
H.R.L.~acknowledges support from a postdoctoral fellowship from the German Research Foundation (DFG) via grant LA 4442/1-1 during the first part of this work. 
Some of the computations were conducted at the Resnick High Performance Computing Center, a facility supported by the Resnick Sustainability Institute at the California Institute of Technology.
The SHCI calculations were performed on the Bridges computer at the Pittsburgh Supercomputing Center under grant PHY170037.
\if\USEACHEMSO1
\end{acknowledgement}
\fi
\new{
\paragraph{Data Availability}
Data for reproducing Figure 2a is available at \url{https://github.com/h-larsson/Cr2Pes22}. Further data is available from the authors upon reasonable request.
}

\if\USEACHEMSO1
\begin{suppinfo}
\else
\section*{Supplementary Information}
\fi
\new{See Supplementary Information (SI) for additional details on the electronic structure methods (Section S1), the vibrational spectrum (S2), the PEC fit (S3), the assignment (S4), the uncertainty estimate of the experimental PEC (S5), as well as references for Figure 1 (S6).}
\if\USEACHEMSO1
\end{suppinfo}
\fi

\providecommand{\latin}[1]{#1}
\makeatletter
\providecommand{\doi}
  {\begingroup\let\do\@makeother\dospecials
  \catcode`\{=1 \catcode`\}=2 \doi@aux}
\providecommand{\doi@aux}[1]{\endgroup\texttt{#1}}
\makeatother
\providecommand*\mcitethebibliography{\thebibliography}
\csname @ifundefined\endcsname{endmcitethebibliography}
  {\let\endmcitethebibliography\endthebibliography}{}

\end{document}